\documentclass[prl,final,twocolumn,showpacs]{revtex4}

\usepackage{graphicx}% Include figure files
\usepackage{dcolumn}% Align table columns on decimal point

\begin{document}

%Title of paper
\title{On the Formation of Copper Linear Atomic Suspended Chains}

\author{F. Sato$^{1}$}
\author{A.S. Moreira$^{2}$}
\author{J. Bettini$^{3}$}
\author{P.Z. Coura$^{4}$}
\author{S.O. Dantas$^{4}$}
\email{dantas@fisica.ufjf.br}
\author{D. Ugarte$^{1,3}$}
\email{ugarte@lnls.br}
\author{D.S. Galv\~{a}o$^{1}$}
\email{galvao@ifi.unicamp.br}

\affiliation{$^{1}$Instituto de F\'{\i}sica Gleb
Wataghin, Universidade Estadual de Campinas, CP 6165, 13083-970
Campinas, SP, Brazil}
\affiliation{$^{2}$Centro Brasileiro de Pesquisas F\'{\i}sicas, R. Dr. Xavier Sigaud 150, 
Rio de Janeiro, 22290-180, RJ, Brazil}
\affiliation{$^{3}$Laborat\'{o}rio Nacional de Luz S\'{\i}ncrotron, CP 6192,
13084-971 Campinas, SP, Brazil}
\affiliation{$^{4}$Departamento de F\'{\i}sica,
ICE, Universidade Federal de Juiz de Fora, 36036-330 Juiz de Fora,
MG, Brazil}

\date{\today}

\begin{abstract}
We report high resolution transmission electron microscopy and classical molecular dynamics 
simulation results of mechanically stretching copper nanowires conducting to linear atomic 
suspended chains (LACs) formation. In contrast with some previous experimental and theoretical 
work in literature that stated that the formation of LACs for copper should not exist our 
results showed the existence of LAC for the [111], [110], and [100] crystallographic 
directions, being thus the sequence of most probable occurence.
\end{abstract}


\maketitle

In the last years an enormous amount of theoretical and experimental effort have
been devoted to the study of nanostructures. The advances of experimental high spatial resolution techniques or microscopies (high resolution transmission electron microscopy (HRTEM), scanning tunneling microscopy (STM), and atomic force microscopy (AFM)) have revealed 
new phenomena at nanoscale.

Among nanostructures, metallic nanowires (NWs) are of great interest due to the observation of very interesting physical phenomena (spin filters, quantized conductance, etc.) and their possible technological application in nanodevices (molecular electronics, nanocontacts, etc.) \cite{revagra,uzi1,ohnishi,sanchez,vrprl,medina,sorensen,zacharias,kang,stafford,legoas,kruger,novaes,legoas2,ugarte}. These structures have displayed quantized conductance of $n$ $2e^{2}/h$ (G$_{0}$), where $n$ is an integer number, $e$ is the electron charge and, $h$ is 
Planck's constant \cite{revagra}. The ultimate NWs are one-atom thick, when suspended linear atomic chains (LACs) are formed by the mechanical stretching.  

Most studies of LACs structures have been focused on Au, but in principle LACs may also possible to form for many other fcc metals (Pt,Ag,Pd,etc.), although we must emphasize that this issue is still rather controversial \cite{bahn,smit,vrbook,vrag,vrmag}. 
In this work, we present a detailed theoretical (molecular dynamics simulations) and experimental (HRTEM) study of mechanical stretching of copper NWs, in particular, addressing the formation of suspended atom chains.

Metallic NWs were produced \textit{in situ} in the HRTEM (JEM-3010 URP 300kV, 0.17 nm point resolution) using the methodology proposed by Kondo and Takayanagi \cite{kontak}. Initially holes are opened at several points in a self-supported metal film by focusing the microscope electron beam ($\sim$120 A/cm$^{2}$). When two holes become very close, nanometric constrictions (bridges) are formed between them. Then, the microscope beam current density is reduced to $\sim$10-30 A/cm$^{2}$ for image acquisition \cite{vrbook}, and the nanowires evolve spontaneously, elongate and finally break, sometimes with the appearance of atomic suspended chains (LACs). These processes are registered using a high sensitive TV camera (Gatan 622SC, 30 frames/s) and a standard video recorder.

The polycrystalline Cu thin films (10-30 nm in thickness) were prepared by thermal evaporation of metal in a standard vacuum evaporator (10$^{-7}$ mbar). A quartz crystal monitor was used to set the evaporation rate of metal source and, subsequently, to measure the equivalent thickness of the film. To prevent oxidation the films were sandwiched between two ($\sim$ 3 nm thick) amorphous carbon layers. Inside the microscope, the carbon layers are removed by electron irradiation \cite{vrbook}.

In Fig. 1, we present a typical HRTEM snapshot showing a LAC composed by two atoms. These represent a direct atomic resolution observation of LAC for copper (see movie 01) \cite{movie}. Due to copper relative low atomic number and smaller lattice parameter, HRTEM image contrast is rather weak, then, the observation of high contrast atomic resolution image is much more difficult to obtain than for other metals such as Au and Pt \cite{vrbook}. Experimentally, LAC formation has been observed for NWs stretched along the [100], [110], and 
[111] crystallographic directions \cite{JuanCu}.

%@@@ Figura 1
\begin{figure}
\includegraphics[width = 6.5 cm]{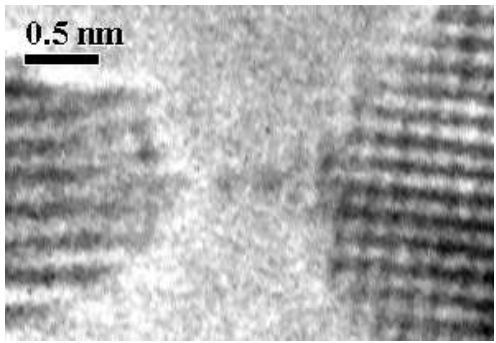}
\caption{Snapshot from HRTEM showing LAC with two hanging and between two grains with different crystallographic orientation along the elongation direction,[110] on left and [100] on right side. Note that the atomic planes of the right apex show the expected d$_{200}$ = 0.18 nm lattice fringes spacing.}
\end{figure}

In order to address the dynamical aspects of the copper LAC formation we have also carried out molecular dynamics simulation using a methodology we have recently developed and, that has been proved to be very effective to address different aspects of metal NW elongation and rupture \cite{coura2004,sato2005,JuanCu}. Our methodology is based on tight-binding molecular dynamics (TB-MD) techniques \cite{CR} using second-moment approximation (SMA) \cite{TAB} with a small set of adjustable parameters.  Details of the methodology were published elsewhere \cite{sato2005}. In this work, the physical stretching that generates the NWs is simulated through structural modification of the wire size (increasing the distance between the outmost layers, for instance). We have considered different elongation rate variations, different temperatures, and initial velocity distributions.

The initial NW configurations consisted of NWs oriented along [111] (hereafter noted [111] NWs), [110], and [100] crystallographic directions with the number of particles varying between 
450-900 copper atoms (typically 5-9 atoms in square basis length and with 10-12 layers along the elongation direction). In order to mimic the bulk structure influence upon NW behavior, we have used buffer atom layers as contacts depending on the crystallographic orientations, two for [110] and [100], and three for [111] in order to preserve the AB(ABC) stacking layer patterns. These buffer layers are frozen and constrained to move only along the stretching directions, while the remaining atoms are free to move into any directions. We have used temperature range of 300-350K, pulling velocities between 0.1-1.5 m/s, time step of 1.0 fs and, atomic interactions up to fifth neighbors. A total of 600 MD simulations were performed (200 for each one of the three analyzed crystallographic directions). In Fig. 2 we present some examples of snapshots from MD simulations for different crystallographic directions. 

Our results have showed the possibility of LAC formation in the final stages of copper NW stretching for all crystallographic studied directions. The number of atoms in the LAC are dependent, as expected, on the pulling velocity, temperature and initial velocity distribution.
As a general trend we have observed that at lower pulling velocity ($\sim$ 0.1m/s) the system has enough time to relax and distribute the internal stress. Then, a better structural order and faceting of the apexes is obtained and, LACs can be well formed as a linear arrangement of hanging atom between to atomically sharp tips. At higher pulling velocity, ($\sim$ 1.5m/s), it is possible to generate longer LACs, but with a poor apex faceting and, an merely approximative linear distribution of suspended atoms along the chain. The forces acting on the LACs atoms just before rupture are about $\sim$1.5nN, what is in excellent agreement with experimentally observed forces for gold NWs \cite{Rubio}. 

%@@@ Figura 2 
\begin{figure}
\includegraphics[width = 8.5 cm]{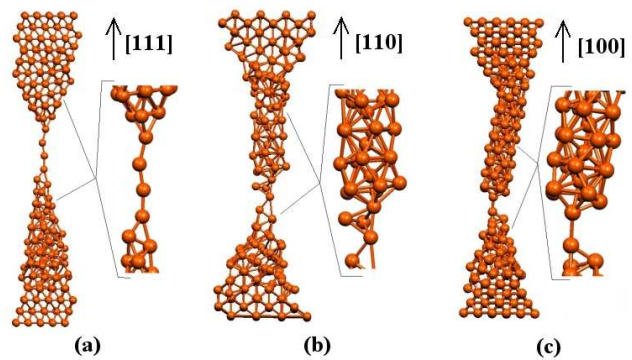}
\caption{Snapshots from MD simulations of Cu NW elongation along (a) [111], (b) [110] and, (c) [100] crystallographic directions, pulling velocity at 1.0m/s, temperature 300K. LAC formation can be clearly identified for the [111] NW in (a). Note that the wires in (b) and (c) exhibit pentagonal structures (see text for discussions)}
\end{figure}

Table 1 summarizes the statistics of LAC formation derived form the MD simulations. Although the formation of suspended chains has been observed for all directions they occur predominantly for [111] and [110] NWs (Figs. 2a,b). Although Cu, Au and Ag share many common structural features, the NWs exhibit clearly differentiated patterns regarding LAC occurence \cite{revagra,vrbook}. Experimentally an theoretically, it has been observed that Au chains are generated with decreasing occurrence rate for [111], [100] and [110] axes \cite{vrprl,coura2004}, while Ag show a completely different pattern ([100], [110] and, [111], in decreasing order \cite{vrag}). Our simulations indicate that, Cu LACs show behavior which is distict from Au and Ag ([111], [110] and, [100] in decreasing order). 
As mentioned above due to intrinsic difficulties associated with copper HRTEM 
experiments, altough we do not have enough experimental data to build a sound statistics, for the available experimental data the general trends are in agreement with our molecular dynamics
results.  

\begin{table}%[H] add [H] placement to break table across pages
\caption{Statistics of LACs formation along crystallographic orientations.}
\begin{ruledtabular}
\begin{tabular}{cccc}
% Lines of table here ending with \\
Crystal Direction & [100] & [110] & [111] \\
Statistics ($\%$) &   1 (2/200)   &   55 (110/200)  &   65 (130/200)  \\
\end{tabular}
\end{ruledtabular}
\end{table}

Non-crystallographic icosahedric structures have been observed for wires along the [110] and [100] directions (Fig. 2b and 2c respectively). These kind of atomic arrangement has been 
recently associated with copper [110] NW in order to explain electric transport experiments displaying a 4.5G$_{0}$ quantum conductance plateaus \cite{JuanCu}. As a consequence of the modified geometrical Wulff construction \cite{marks}, it can expected that [110] NWs form icosahedric structures with five fold axes symmetry \cite{JuanCu}. IN order to explain why pentagonal wires may be formed for Cu [100] NWs, we must also consider that the structural rearrangement in the apexes, may sometimes generate quite asymmetric junctions. This may lead to the formation of NWs, which may be tilted in relation to the original direction; in fact, the NW in Fig. 2c is tilted by $\sim$35 degrees from the [100] axis. Rodrigues et al \cite{vrprl} have already revealed that the atomic arrangement of atomic-size NWs adjusts such that one of the main gold zone axes lies approximately parallel to the elongation direction, independent of the apex crystal orientation, and this structural adjustment occurs mainly
concentrated around the narrowest neck region. Then, it is not surprising that a Cu [100] NW  tilted by $\sim$35 degress (shown in Fig. 2C) could adopt a [110]-like atomic structure (icosahedric), what in fact means a minor 10 degrees disorientation from a [110] axis.  This is quite well reproduced in Fig. 2c, and it is a good indication that the simulations are confirming expected structures. 

As discussed above, dynamical features of the neck or apex atomic reorganization influence LAC formation and, then structural/morphological aspects are very important. In this sense the size of crystalline grains (or initial configuration for a simulation) might play an significant role. In order to evaluate the importance of the aspect ratio of the initial NWconfiguration, we have carried out further simulations, where we systematically increased the lateral dimensions. We  have observed that after certain limit, we start to observe the appearance of parallel columnar NWs; in Fig. 3 we present an example of this phenomena. The general trends of each constriction are basically the same as in the case of only one collum.

%@@@ Figura 3
\begin{figure}
\includegraphics[width = 6 cm]{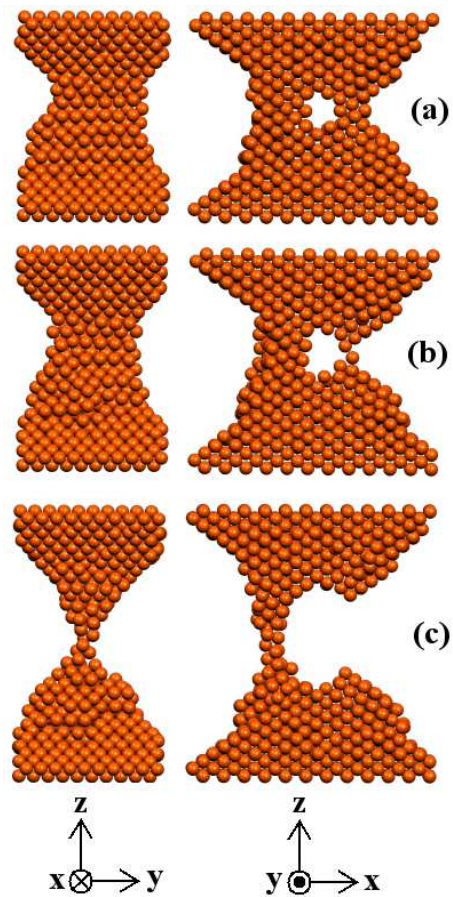}
\caption{Sequential snapshots (a), (b) and (c) of MD simulations showing the occurence of multi-collumnar structures in a system with 1800 copper atoms with pulling velocity (4.0m/s alloy y axis) direction along [111] crystallographic direction, temperature  350K. Left side pictures are view along x axis and right side pictures along z axis. These multi-columnar 
structures could form LAC one- or two-atom-long.} 
\end{figure}

The existence of this kind of structures was first proposed by Marqu\'{e}z and Garcia \cite{Correia}. Although in our HRTEM experiments they are unlike to occur due to fabrication procedures (thinning a bridge between two growthing holes). They are a real possibility in mechanically controllable break junction (MCBJ) experiments \cite{revagra}.

In summary our HRTEM and molecular dynamics simulatiosn have shown that, in contrast
with some previous theoretical and experimental predictions,  copper suspended linear atomic
chains are possible for [111], [110], and [100], being this the sequence of likely
ocurrence.  

This work was supported by LNLS, CNPq, FAPESP, IMMP/MCT, IN/MCT and CAPES. The authors acknowledge the invaluable help of the LNLS staff, in particular P.C. Silva for sample preparation. The authors also wish to thank J. C. Gonzalez, L. G. C. Rego, and A. R. Rocha for helpful discussions.

\end{document}